\documentclass[twocolumn,prl,superscriptaddress,showpacs]{revtex4}

\usepackage{graphicx}
\usepackage{dcolumn}
\usepackage{multirow}
\usepackage{bm}
\usepackage{times}
\usepackage{color}
\usepackage{amstext}
\usepackage{amsmath}
\usepackage{amsfonts}
\usepackage[bookmarks=true,colorlinks=true,pdfpagemode=UseNone,pdfstartview=FitH,linkcolor=blue,urlcolor=blue,citecolor=blue]{hyperref}
\usepackage{ocg,dashbox,color,fancyhdr}%
\begin{document}

\title{Eliashberg approach to superconductivity-induced infrared anomalies in $\bf\textrm{Ba}_{0.68}\textrm{K}_{0.32}\textrm{Fe}_2\textrm{As}_2$}

\author{A. Charnukha}
\author{O. V. Dolgov}
\affiliation{Max-Planck-Institut f\"ur Festk\"orperforschung, Heisenbergstrasse 1, D-70569 Stuttgart, Germany}
\author{A. A. Golubov}
\affiliation{Faculty of Science and Technology and MESA+ Institute of Nanotechnology, 7500 AE Enschede, The Netherlands}
\author{Y. Matiks}
\author{D. L. Sun}
\author{C. T. Lin}
\author{B. Keimer}
\affiliation{Max-Planck-Institut f\"ur Festk\"orperforschung, Heisenbergstrasse 1, D-70569 Stuttgart, Germany}
\author{A. V. Boris}
\affiliation{Max-Planck-Institut f\"ur Festk\"orperforschung, Heisenbergstrasse 1, D-70569 Stuttgart, Germany}

\begin{abstract}
We report the full complex dielectric function of high-purity $\textrm{Ba}_{0.68}\textrm{K}_{0.32}\textrm{Fe}_2\textrm{As}_2$ single crystals with $T_{\mathrm{c}}=38.5\ \textrm{K}$ determined by wide-band spectroscopic ellipsometry at temperatures $10\leq T\leq300\ \textrm{K}$. We discuss the microscopic origin of superconductivity-induced infrared optical anomalies in the framework of a multiband Eliashberg theory with two distinct superconducting gap energies $2\Delta_{\mathrm{A}}\approx6\ k_{\mathrm{B}}T_{\mathrm{c}}$ and $2\Delta_{\mathrm{B}}\approx2.2\ k_{\mathrm{B}}T_{\mathrm{c}}$. The observed unusual suppression of the optical conductivity in the superconducting state at energies up to $14\ k_{\mathrm{B}}T_{\mathrm{c}}$ can be ascribed to spin-fluctuation--assisted processes in the clean limit of the strong-coupling regime.
\end{abstract}


\maketitle

\par The discovery of iron-based superconductors~\cite{kamihara} has generated significant experimental and theoretical effort to unravel the mechanism of high-temperature superconductivity in these compounds. This effort has yielded a comprehensive experimental description of the electronic structure at the Fermi level, which includes multiple Fermi surface sheets in a good agreement with density functional calculations~\cite{Paglione_review_2010,Johnston_Review_2010}. Partial nesting between at least two of these sheets leads to a spin-density-wave instability that renders the metallic parent compounds antiferromagnetic. In the superconducting compounds spin fluctuations become the source of strong repulsive interband interactions and might give rise to superconductivity with different signs on these sheets~\cite{Mazin_NatureInsights_2010}.

The most incisive experimental data have been obtained on high-quality single-crystals of iron pnictides with the so-called 122 structure, for instance $\textrm{BaFe}_2\textrm{As}_2$, with K substituted for Ba or Co for Fe, resulting in hole and electron doping, respectively. In all of these materials five Fermi surface sheets have been identified in calculations and confirmed by numerous independent experimental studies~\cite{Paglione_review_2010,Johnston_Review_2010}: in the reduced Brillouin-zone scheme these are three hole pockets at the $\Gamma$ point and two almost degenerate electron pockets at the $X$ point with nesting between hole and electron sheets. Among all of these $122$ materials, the optimally hole-doped compound $\textrm{Ba}_{0.68}\textrm{K}_{0.32}\textrm{Fe}_2\textrm{As}_2$ (BKFA) has the highest transition temperature of $38.5\ \textrm{K}$. Due to their exceptional quality, crystals of this compound are well suited as a testbed for theoretical models. A four-band Eliashberg theory with strong interband couping has already proven successful in accounting for the transition temperature, as well as the temperature dependence of the free energy and superconducting gaps of this compound~\cite{PhysRevLett.105.027003}. This analysis has made clear that a satisfactory description of the bulk thermodynamical properties in the superconducitng state can only be obtained via strong coupling to spin fluctuations or other bosons with spectral weight below $50\ \textrm{meV}$. 

\begin{figure}[tb]
\includegraphics[width=3.4in]{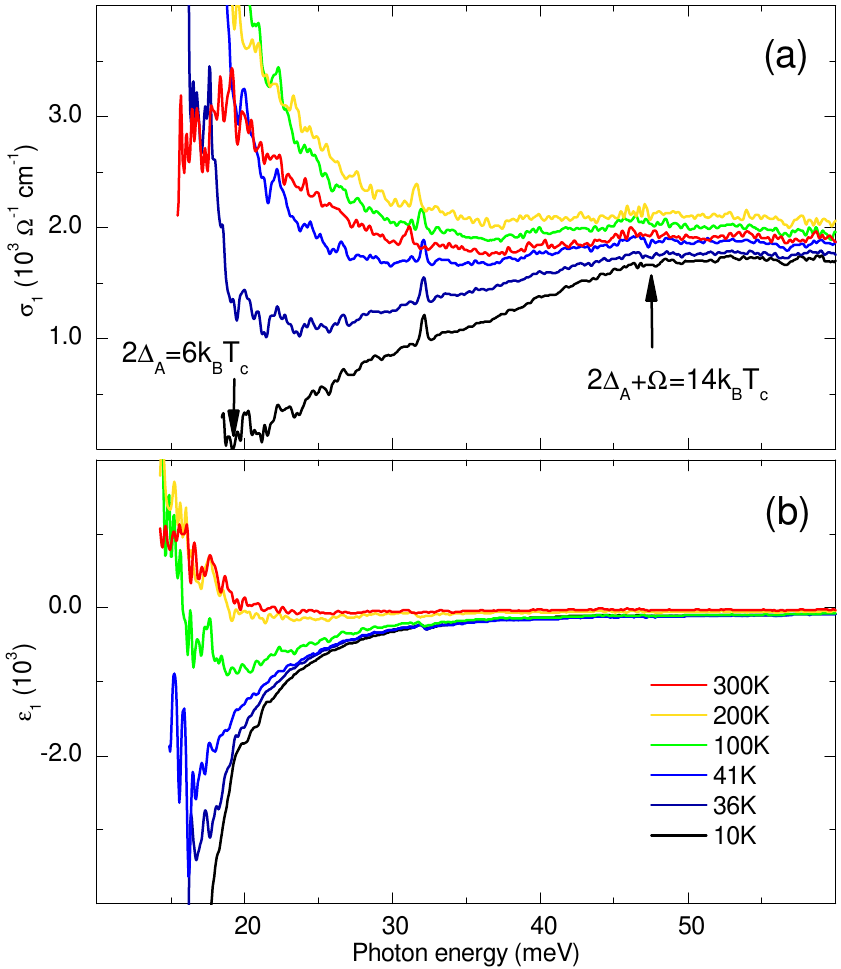}
\caption{\label{fig:firexperiment} Real part of the (a) optical conductivity and (b) dielectric function in the far-infrared spectral region. Two characteristic superconductivity energy scales are present: $6\ k_{\mathrm{B}}T_{\mathrm{c}}$ and $14\ k_{\mathrm{B}}T_{\mathrm{c}}$.}
\end{figure}
In this Letter we extend the approach of Ref.~\onlinecite{PhysRevLett.105.027003} to describe the far-infrared properties of the same BKFA single crystal. We show that major characteristic features of superconductivity can be explained within a strong-coupling Eliashberg approach with two distinct values of the superconducting energy gap $2\Delta_{\mathrm{A}}\approx6\ k_{\mathrm{B}}T_{\mathrm{c}}$ and $2\Delta_{\mathrm{B}}\approx2.2\ k_{\mathrm{B}}T_{\mathrm{c}}$, in quantitative agreement with angle-resolved photoemission~\cite{PhysRevB.79.054517,PhysRevLett.105.117003,PhysRevB.83.020501}, scanning-tunneling microscopy~\cite{Shan-Wen_STM_2011} and specific-heat measurements~\cite{PhysRevLett.105.027003}. We also demonstrate that within this approach the qualitative differences in the infrared spectra of electron- and hole-doped 122 compounds are reproduced by strong-coupling calculations in the clean and dirty limits (weak and strong impurity scattering), respectively.

The optimally-doped BKFA single crystals were grown in zirconia crucibles sealed in quartz ampoules under argon atmosphere~\cite{Lin_BKFA_growth_2010}. From
DC resistivity, magnetization and specific-heat measurements we obtained $T_{\mathrm{c}}=38.5\pm0.2\ \textrm{K}$. The sample surface was cleaved prior to every optical measurement. The full complex dielectric function $\varepsilon(\omega)$ was obtained in the range $0.01-6.5\ \textrm{eV}$ using broadband ellipsometry, as described in Ref.~\cite{boris:027001}. In this work we focus on the itinerant charge carrier response contained within the far-infrared spectral range measured at the infrared beamline of the ANKA synchrotron light source at Karlsruhe Institute of Technology, Germany. The contribution of the interband transitions has been eliminated based on a dispersion analysis in the entire spectral range~\cite{supplementary.material.prl}. 
\begin{figure}[tb]
\includegraphics[width=3.4in]{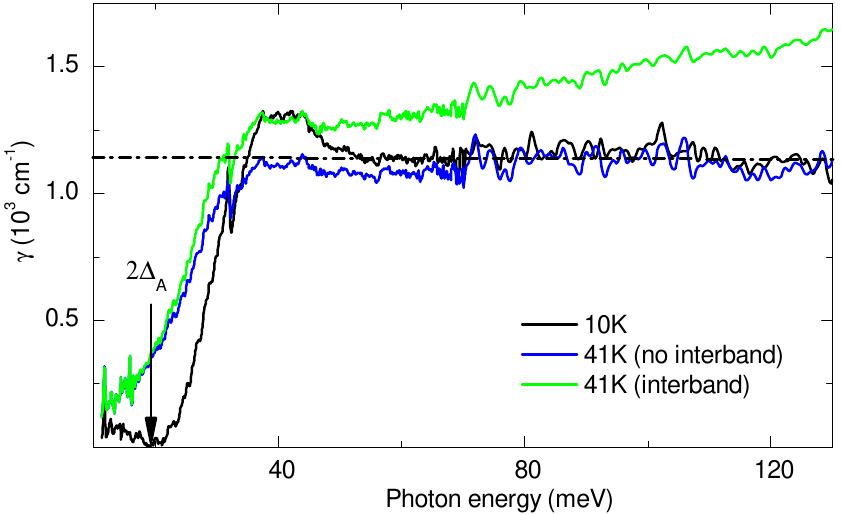}
\caption{\label{fig:scattering} Optical scattering rate obtained from the experimental data at $41$ and $10\ \textrm{K}$ within the extended-Drude model, with the contribution of the interband transitions subtracted (blue and black lines, respectively) and at $41\ \textrm{K}$ without subtraction (green line). Dash-dotted line indicates the saturation level of the high-energy optical scattering rate.}
\end{figure}
\par The optical response of BKFA in the far-infrared spectral range is shown in Fig.~\ref{fig:firexperiment}(a, b) respectively for the real parts of optical conductivity $\sigma(\omega)=\sigma_1(\omega)+i\sigma_2(\omega)$ and dielectric function $\varepsilon(\omega)=1+4\pi i\sigma(\omega)/\omega$. It is dominated by the contribution of the itinerant charge carriers manifested in negative values of $\varepsilon_1(\omega)$. Figure~\ref{fig:firexperiment}(a) also reveals a peak around $15-20\ \textrm{meV}$ in $\sigma_1(\omega)$ with a concomitant upturn in $\varepsilon_1(\omega)$ at higher temperatures indicating the presence of a collective excitation. The superconducting transition is accompanied by the suppression of the optical conductivity up to $50\ \textrm{meV}$~($14k_{\mathrm{B}}T_{\mathrm{c}}$). The corresponding missing area in $\sigma_1(\omega)$ between $41$~and $10\ \textrm{K}$,  $\int_{0^+}^{6\Delta_{\mathrm{A}}}\Delta\sigma_1(\omega)d\omega=(8\lambda_{\mathrm{L}}^2)^{-1}$, manifests itself as a characteristic $-1/(\lambda_{\mathrm{L}}\omega)^2$ contribution to $\varepsilon_1(\omega)$ in Fig.~\ref{fig:firexperiment}(b) in the superconducting state. The London penetration depth $\lambda_{\mathrm{L}}=2200$~\AA\ extracted from these data is consistent with other measurements~\cite{PhysRevLett.101.107004}. At energies close to the optical superconducting gap $2\Delta_{\mathrm{A}}\approx20\ \textrm{meV}$ one of the directly measured ellipsometric angles $\Psi(\omega)$ approaches its critical value of $45^\circ$ at the superconducting transition, which implies that the reflectivity of the sample approaches unity and its optical conductivity $\sigma_1(\omega)$ is close to zero. Remarkably, Fig.~\ref{fig:firexperiment}(a) shows a quasilinear dependence of $\sigma_1(\omega)$ in the superconducting state from $2\Delta_{\mathrm{A}}$ to as high as $14k_{\mathrm{B}}T_{\mathrm{c}}$, in a stark contrast to the electron-doped 122 compounds~\cite{PhysRevB.82.174509,PhysRevB.81.214508,PhysRevB.82.100506}. In the latter the optical conductivity at $2\Delta_{\mathrm{A}}$ decreases abruptly upon cooling below $T_{\mathrm{c}}$, but only a weak superconductivity-induced modification is observed at higher energies. The quasilinear behavior in BKFA cannot be reconciled with the widely used for pnictides Mattis-Bardeen theory~\cite{Zimmermann199199}, a weak-coupling extension of the BCS theory to finite impurity scattering. As all optimally-doped 122 pnictide superconductors appear to be in the strong-coupling regime, the Eliashberg theory~\cite{PhysRev.156.470} has to be used in order to obtain an adequate description of the optical properties.
\begin{figure}[!b]
\includegraphics[width=3.4in]{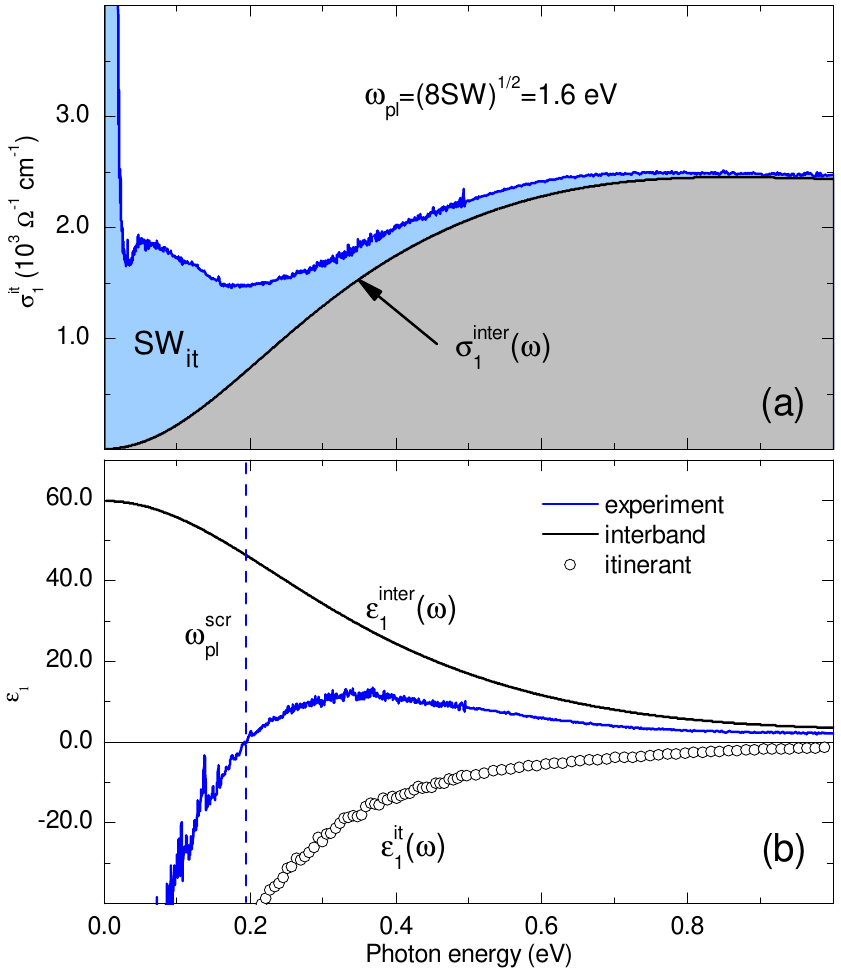}
\caption{\label{fig:zerocrossing} (a)~Real part of the optical conductivity at 41K (blue line). The contribution of itinerant charge carriers (blue area) is obtained by subtracting all interband transitions $\sigma^{\mathrm{inter}}_1(\omega)$ (gray area) from the optical response. (b)~Real part of the dielectric function at 41K (blue line). The free-charge-carrier response $\varepsilon^{\mathrm{it}}_1(\omega)$ (open circles) is obtained by eliminating all interband transitions $\varepsilon^{\mathrm{inter}}_1(\omega)$ (black solid line). The blue dashed line indicates the screened plasma frequency at 41K.}
\end{figure}

Signatures of a boson pairing mediator of the Eliashberg theory come from a qualitative analysis of the optical conductivity within the extended Drude model. It implies that the optical scattering rate is related to the far-infrared optical response as $\gamma(\omega)=\textrm{Re}[\omega_{\mathrm{pl}}^2/4\pi\sigma^{\mathrm{it}}(\omega)]$, where the superscript 'it` stands for 'itinerant` and implies that the contribution of all interband transitions {\it must} be subtracted from the experimentally obtained optical conductivity. The optical scattering rate shows clear evidence of an intermediate boson irrespective of complications due to the multiband character of the compound. Figure~\ref{fig:scattering} plots $\gamma(\omega)$ of BKFA at $41\ \textrm{K}$~(blue line) and $10\ \textrm{K}$~(black line) for $\omega_{\mathrm{pl}}=[8\textrm{SW}_{\mathrm{it}}]^{1/2}=[8\int_{0}^{\infty}\sigma_1^{\mathrm{it}}(\omega)d\omega]^{1/2}=1.6\ \textrm{eV}$ with all interband transitions subtracted in both cases ($\textrm{SW}_{\mathrm{it}}$ corresponds to the blue shaded area in Fig.~\ref{fig:zerocrossing}(a)). In the superconducting state, no scattering is expected up to photon energies exceeding the binding energy of the Cooper pairs. Thus the onset of the optical scattering rate marks the optical energy gap $2\Delta_{\mathrm{A}}=20\ \textrm{meV}$. Saturation of $\gamma(\omega>50\ \textrm{meV})$ at $1100\ \textrm{cm}^{-1}$ indicates that the boson spectral function is contained well below $50\ \textrm{meV}$~\cite{Shulga1991266}. It is important to emphasize that, due to the multiband character of the iron pnictides, an analysis of the optical scattering rate in the framework of a single-band Eliashberg theory is potentially misleading. Moreover, also shown in Fig.~\ref{fig:scattering} is a spectrum that directly results from the experimental data, without accounting for the interband transitions. It becomes clear that an increase in the scattering rate at higher energies that might be ascribed to strong electron correlations can result from an unsubtracted contribution of the interband transitions to the complex optical conductivity. This is especially important in iron pnictides since the lowest lying interband transition at about $0.5\ \textrm{eV}$ contributes to an anomalously large value of the low-energy dielectric permittivity $\varepsilon_\infty$~\cite{footinbib_fir2011} due to the high polarizability of the Fe-As bonds~\cite{2010arXiv1009.5915C}. In order to reconcile the bare plasma frequency of $1.6\ \textrm{eV}$ (see Fig.~\ref{fig:zerocrossing}(a)) with the zero-crossing in $\varepsilon_1(\omega)$ at $0.2\ \textrm{eV}$ (blue line in Fig.~\ref{fig:zerocrossing}(b)) $\varepsilon_\infty$ has to be as large as 60, consistent with the contribution of the interband transitions $\varepsilon^{\mathrm{inter}}_1(\omega)$ determined by means of the dispersion analysis, as shown in Fig.~\ref{fig:zerocrossing}(b). Such $\varepsilon_\infty$ is thus an order of magnitude larger than in any other high-temperature superconductor (e.~g.~$\approx5$ in cuprates~\cite{A.V.Boris04302004}). Recently, a similarly high value in a conventional superconductor was inferred from reflectivity measurements on elementary bismuth~\cite{PhysRevLett.104.237401}.

To determine the microscopic origin of the high-energy anomaly $2\Delta_{\mathrm{A}}<\hbar\omega<14k_{\mathrm{B}}T_{\mathrm{c}}$ in the real part of the optical conductivity in Fig.~\ref{fig:firexperiment}(a) we use a four-band Eliashberg theory that proved successful in explaining thermodynamical data obtained on the same compound~\cite{PhysRevLett.105.027003}. The $4\times4$ matrix of coupling constants and 4 densities of states characterizing this model are highly constrained by thermodynamic, transport and photoemission data~\cite{PhysRevLett.105.027003,PhysRevB.79.054517,PhysRevLett.105.117003,PhysRevB.83.020501,Shan-Wen_STM_2011}, and the same set of parameters is used here. In principle, an additional set of 4 plasma frequencies and a $4\times4$ matrix of intraband/interband impurity scattering rates has to be taken into account to describe the optical response. However, this parameter set can be strongly reduced based on the following considerations.

A substantial simplification is made possible by a projection of the four-band model onto an effective two band model motivated by the observation of two distinct groups of superconducting energy gaps in a variety of experiments~\cite{PhysRevLett.105.027003,PhysRevB.79.054517,PhysRevLett.105.117003,PhysRevB.83.020501,Shan-Wen_STM_2011}. These gaps can be identified as a single gap $\Delta_B$ on the outer hole-like Fermi surface and a group of three gaps of magnitude $\sim\Delta_A$ on the inner hole-like and the two electron-like Fermi surfaces. Minimizing the ground-state energy subject to this grouping constraint yields an effective two-band model~\cite{supplementary.material.prl}. Furthermore, as the superconducting transition temperature of BKFA appears to be only weakly correlated with the residual resistivity (which is a measure of the impurity scattering), off-diagonal elements of the impurity scattering matrix can be neglected (see Table~S1 in Ref.~\onlinecite{supplementary.material.prl}).

Given the boson spectrum centered at $13\ \textrm{meV}$ (see supplementary online material in Ref.~\onlinecite{PhysRevLett.105.027003}) consistent with the energy of the spin resonance excitation in this compound~\cite{Osborn_INS_BKFA_2008,footinbib_SF2011} one obtains the following two-band model coupling matrix: $\lambda_{\mathrm{AA}}=4.36,\ \lambda_{\mathrm{BB}}=0.2,\ \lambda_{\mathrm{AB}}=-0.35,\ \lambda_{\mathrm{BA}}=-0.5$, with the fractional density of states being $N_{\mathrm{A}}/(N_{\mathrm{A}}+N_{\mathrm{B}})=0.59$~\cite{supplementary.material.prl}. The first effective intraband coupling constant is an order of magnitude larger than predicted for the intraband electron-phonon coupling~\cite{boeri:026403}. It does not, however, bear any physical meaning by itself but rather incorporates contributions from {\it three} different bands. We reiterate that the coupling matrix has been inferred from prior measurements. In this way, only two intraband impurity scattering rates enter as free parameters of the theory in addition to the plasma frequencies of the bands.

\begin{figure}[tb]
\includegraphics[width=3.4in]{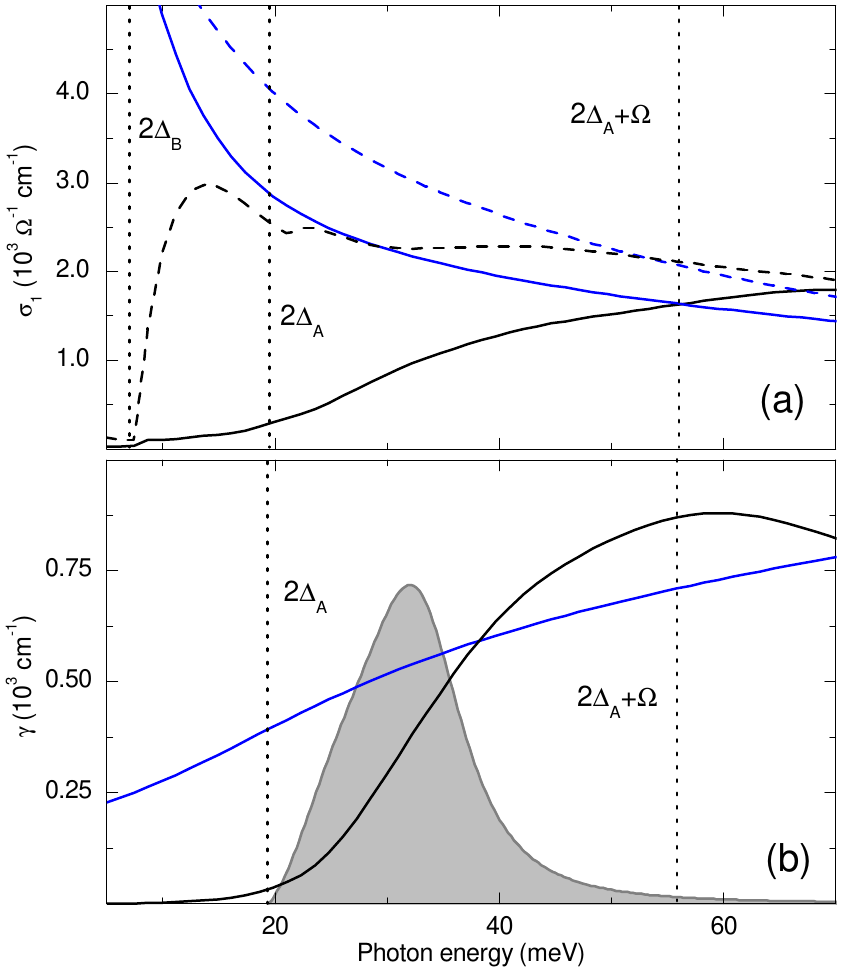}
\caption{\label{fig:eliashberg_sigma} (a)~Real part of the far-infrared conductivity obtained within the two-band Eliashberg theory~(see text) at $40\ \textrm{K}$ (blue lines) and $10\ \textrm{K}$ (black lines) in the clean limit $\gamma_{\mathrm{A}}=\gamma_{\mathrm{B}}=1\ \textrm{cm}^{-1}$ (solid lines) and dirty limit $\gamma_{\mathrm{A}}=\gamma_{\mathrm{B}}=200\ \textrm{cm}^{-1}$ (dashed lines).~(b)~Optical scattering rate in the clean limit from the same model. The gray area shows the normalized boson spectral function $B(\omega)$ used in the calculation, displaced from zero by $2\Delta_{\mathrm{A}}$ to assist interpretation in the superconducting state~\cite{footinbib_SF2011}.}
\end{figure}
In our calculation we consider two clean bands with $\gamma_{\mathrm{A}}=\gamma_{\mathrm{B}}=1\ \textrm{cm}^{-1}$. As the bare plasma frequencies of all bands are similar it follows that the spectral weight of band~$\mathrm{A}$ has to be much larger than that of band~$\mathrm{B}$. Assigning $80\%$ of the spectral weight to the effective band we obtain the results presented as solid lines in Fig.~\ref{fig:eliashberg_sigma}(a). The high-energy anomaly at $14k_{\mathrm{B}}T_{\mathrm{c}}$ is naturally captured by the model without resorting to additional gaps. Its energy is given by $2\Delta_{\mathrm{A}}+\Omega$, where $\Omega$ is the characteristic frequency of the boson spectrum, as shown in Fig.~\ref{fig:eliashberg_sigma}(b) (gray shaded area, displaced from zero by $2\Delta_{\mathrm{A}}$). This calculation also accounts for the fact that only the biggest superconducting gap is visible in the optical responce of BKFA due to a small contribution of band~B ($20\%$ of the spectral weight) to the overall optical conductivity. This leads to two possible levels of the impurity scattering rate of band~B, which has to be either very small $\gamma_{\mathrm{B}}\approx1\ \textrm{cm}^{-1}$ or very large at about $1000\ \textrm{cm}^{-1}$. The latter value provides a better description of the optical scattering rate~(see interactive simulation in~\cite{supplementary.material.prl}) and DC transport~\cite{2010arXiv1011.1900G}. However, such a large disparity between the charge carriers is hard to reconcile with the Hall and de~Haas--van~Alphen experiments, which imply that the impurity scattering rate of the holes is no more than one order of magnitude higher than that of the electrons~\cite{PhysRevB.80.140508,PhysRevLett.101.216402}. This residual uncertainty notwithstanding, our results show that the impurity scattering rate of band~A must be very small, because the energy $2\Delta_{\mathrm{A}}+\Omega$ is no longer discernible in the simulated spectra when $\gamma_{\mathrm{A}}$ increases~(interactive simulation in~\cite{supplementary.material.prl}). The region of linear increase of $\sigma_1(\omega)$ is related to the linear segment in the boson spectrum and can only be observed in a very clean material.

The same reduced two-band model can be applied to the case of $\textrm{BaFe}_{1.85}\textrm{Co}_{0.15}\textrm{As}_2$ (BFCA). In this compound the spin resonance excitation occurs at a very similar energy of $10\ \textrm{meV}$~\cite{Inosov_BFCA_2009}. A boson spectrum centered at this energy is also consistent with Andreev-reflection measurements~\cite{PhysRevLett.105.237002}. Recently, a comprehensive specific-heat study of this compound at different Co-doping levels has been carried out~\cite{0295-5075-91-4-47008}. The analysis of the experimental data in the framework of the two-band $\alpha$-model indicates that the largest gap develops in the band with the largest electronic density of states, providing further evidence that several bands contribute to the strongly-coupled band in the reduced two-band model. Figure~\ref{fig:eliashberg_sigma}(a) (dashed lines) shows that a calculation within the same reduced two-band model qualitatively reproduces the far-infrared optical conductivity of BFCA~\cite{PhysRevB.82.174509,PhysRevB.81.214508,PhysRevB.82.100506} when both bands are assumed to be dirty with $\gamma_{\mathrm{A}}=\gamma_{\mathrm{B}}=200\ \textrm{cm}^{-1}$ and a redistribution of the spectral weight between the bands is taken into account as $\omega^2_{\mathrm{pl,A}}\approx\omega^2_{\mathrm{pl,B}}$. The model captures the two prominent superconductivity-induced anomalies clearly observed in experiments: the steep onset of absorption at the value of the small gap $2\Delta_{\mathrm{B}}$ and the weaker superconductivity-induced changes of the optical conductivity extending up to $18k_{\mathrm{B}}T_{\mathrm{c}}$. The redistribution of the spectral weight between the bands in BFCA compared to BKFA implied by our analysis is justified by doping with different carriers in the two compounds, while the large difference in the their impurity scattering rates is a natural consequence of the difference in doping mechanisms by chemical substitution, which directly affects the FeAs layers in BFCA, but not in BKFA.

In summary, a qualitative description of superconductivity-induced optical anomalies in the far-infrared optical conductivity of $\textrm{Ba}_{0.68}\textrm{K}_{0.32}\textrm{Fe}_2\textrm{As}_2$ is obtained in the framework of an effective two-band Eliashberg theory with a strong coupling to spin fluctuations reduced from its four-band counterpart. The linear increase of absorption above the larger superconducting gap can only be observed when the effective band is extremely clean. The same model in the dirty limit provides a good qualitative explanation of the optical conductivity of the optimally electron-doped BFCA consistently in the strong-coupling regime.

This project was supported by the German Science Foundation under grant BO 3537/1-1 within SPP 1458. We gratefully acknowledge Y.-L. Mathis for support at the infrared beamline of the synchrotron facility ANKA at the Karlsruhe Institute of Technology and P.~Popovich for taking part in some of the measurements.


\newpage
\setcounter{figure}{0}
\renewcommand\thefigure{S\arabic{figure}}
\renewcommand\theequation{S\arabic{equation}}
\renewcommand\thetable{S\arabic{table}}
\renewcommand{\cite}[1]{[S\citenum{#1}]}
\renewcommand{\bibnumfmt}[1]{S#1.}
\onecolumngrid
\begin{center}
{\large\bf Supplementary online material for the article\\Eliashberg approach to superconductivity-induced infrared anomalies in $\bf\textrm{Ba}_{0.68}\textrm{K}_{0.32}\textrm{Fe}_2\textrm{As}_2$}
\vskip0.17in
A.~Charnukha$^1$, O.~V.~Dolgov$^1$, A.~A.~Golubov$^2$, Y.~Matiks$^1$, D.~L.~Sun$^1$, C.~T.~Lin$^1$, B.~Keimer$^1$, and A.~V.~Boris$^1$
\vskip0.05in
{\small\it$^\mathit{1}$Max-Planck-Institut f\"ur Festk\"orperforschung, Heisenbergstrasse 1, D-70569 Stuttgart, Germany}\\
{\small\it$^\mathit{2}$Faculty of Science and Technology and MESA+ Institute of Nanotechnology, 7500 AE Enschede, The Netherlands}
\end{center}
\twocolumngrid
\section{Dispersion analysis}
\begin{figure}[!t]
\includegraphics[width=\columnwidth]{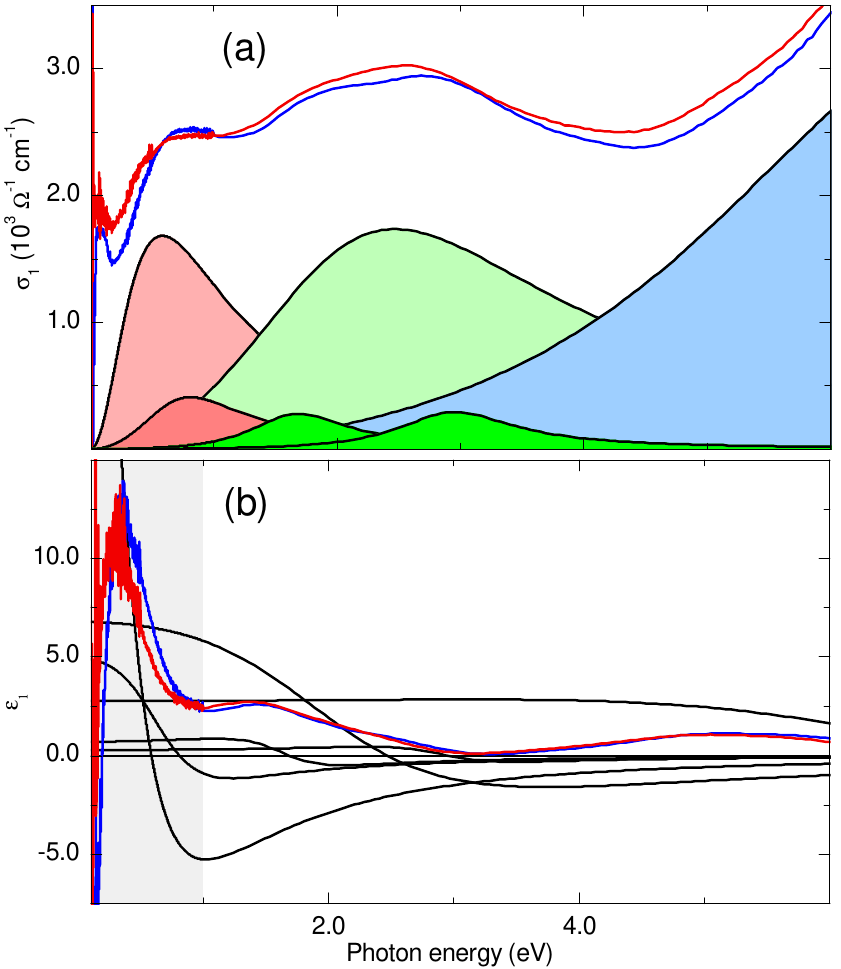}
\caption{\label{fig:dispanalysis}Real part of the (a)~optical conductivity and (b)~dielectric function~(b) at 300K (red line) and 10K (blue line). Interband transitions inferred from the dispersion analysis (gray lines). The shaded region is enlarged in Fig.~3(b) of the main text.}
\end{figure}

The full complex dielectric function $\varepsilon(\omega)=\varepsilon_1(\omega)+i\varepsilon_2(\omega)$ obtained experimentally in the range from $12\ \textrm{meV}$~to~$6.7\ \textrm{eV}$ was analyzed in the Drude-Lorentz model:%
\begin{equation}%
\varepsilon(\omega)=1-\frac{\omega_{\mathrm{pl}}^2}{\omega^2+i\gamma\omega}+\sum_{j=1}^{n}\frac{\Delta\varepsilon_j\omega_{0j}^2}{(\omega_{0j}^2-\omega^2)-i\Gamma_j\omega}\nonumber,
\end{equation}where $(\omega_{\mathrm{pl}},\ \gamma)$ are the plasma frequency renormalized by interaction with the mediating boson and optical scattering rate, and $(\Delta\varepsilon_j,\omega_{0j},\ \Gamma_j)$ are the DC permittivity contribution, center frequency and the width of the Lorentzian oscillators used to model the interband transitions, respectively. The results of this analysis are presented in Fig.~\ref{fig:dispanalysis} for $10\ \textrm{K}$ (blue line - experimental data, black lines - separate Lorentz contributions). To display the scale of the temperature-induced variation of the interband transitions the experimental spectrum at $300\ \textrm{K}$ is also shown (red line). The lowest interband transition in this material lies around $0.5\ \textrm{eV}$ and significantly contributes to the AC polarizability of the system, as is evident from Fig.~3(b) of the main text. The residual optical response (open circles and inset in Fig.~3(b)) was studied after the subtraction of all thus determined interband transitions down to $0.5\ \textrm{eV}$. An unscreened bare plasma frequency of $1.6\ \textrm{eV}$ at $41\ \textrm{K}$ was consistently obtained from the spectral weight of the residual response $\textrm{SW}=\int_{0}^{\infty}\sigma_1^{\mathrm{it}}(\omega)d\omega$ as $\omega_{\mathrm{pl}}=\sqrt{8\textrm{SW}}$ and a simultaneous fit of the real and imaginary parts of the dielectric function at high energies.
\section{Effective two-band model for pnictides}
\subsection{Multiband Eliashberg formalism}
The Eliashberg theory of superconductivity~[S\onlinecite{PhysRev.156.470s},S\onlinecite{PhysRev.156.487s}] extended to the multiband case~\cite{PhysRevB.72.024504} has already proven successful in describing the thermodynamical properties of iron pnictides~\cite{PhysRevLett.105.027003s}. Here we employed the same formalism to account for the far-infrared optical response of these compounds. The main ingredient of the theory is the total spectral function
of the electron-boson interaction $B(\omega)$ (Eliashberg function; analogous to that of the electron-phonon interaction $\alpha ^{2}F(\omega)$). In a four band system it can be decomposed into 16 functions $B(\omega) _{ij}$, where $i$ and $j$
label the four Fermi surface sheets ($i,j=1,2,3,4$). The \textit{standard} Eliashberg functions determine the superconducting and thermodynamical properties such as the superconducting transition temperature and gaps, electronic specifc heat, de~Haas-van~Alphen mass renormalizations etc. and are defined as 
\[
B(\omega)_{ij}=\frac{1}{N_{i}}\sum_{{\bf k,k}%
^{\prime },\nu }\left| g_{{\bf k,k}^{\prime }}^{ij,\nu }\right| ^{2}\delta
(\varepsilon _{{\bf k}}^{i})\delta (\varepsilon _{{\bf k^{\prime }}%
}^{j})\delta (\omega -\omega _{{\bf k-k^{\prime }}}^{\nu }),\nonumber
\]
where $N_{i}$ is the partial density of states per spin on the $i$'th sheet of the Fermi surface, and $g_{{\bf k,k}^{\prime
}}^{ij}$ is the matrix element of electron-boson interactions.

Transport and electrodynamical properties are defined by 16 {\it %
transport} Eliashberg functions (which enter the Boltzmann kinetic equation)
\begin{eqnarray}
B(\omega)_{tr\;ij}^{\alpha\beta}&=&\frac{1%
}{2N_{i}\left\langle v_{Fi}^{\alpha \text{ }2}\right\rangle }\sum_{{\bf %
k,k}^{\prime },\nu }\left| g_{{\bf k,k}^{\prime }}^{i,j,\nu }\right|
^{2}
\nonumber\\
&&\hskip-25pt\times(v_{Fi}^{\alpha }({\bf k})-v_{Fj}^{\beta }({\bf k}^{\prime }))^{2}\delta (\varepsilon _{{\bf k}}^{i})\delta (\varepsilon _{{\bf k}%
^{\prime }}^{j})\delta (\omega -\omega _{{\bf k-k}^{\prime }}^{\nu }),
\nonumber
\end{eqnarray}
where $v_{Fi}^{\alpha }$ is the $\alpha $-th Cartesian component of the Fermi
velocity on Fermi surface $i$. The average Fermi velocity is related to the
plasma frequency by the standard expression $\omega _{pl\;i}^{2}=8\pi e^{2}N_{i}\left\langle v_{Fi}^{2}\right\rangle
=8\pi e^{2}\sum_{{\bf k}}v_{Fi}^{2}({\bf k})\delta (\varepsilon _{{\bf k}%
}^{i}).$
All Eliashberg functions satisfy the symmetry relations $M_{i}B _{ij}=M_{j}B _{ji},$ where $M_{i}=N_{i}$ and $%
M_{i}=\omega _{pl\,i}^{2}$ for the standard and transport Eliashberg functions, respectively.

\subsection{Role of impurities and defects}

Both normal and superconducting properties of a multiband superconductor significantly depend on impurity scattering. Unlike in conventional superconductivity, one has to distinguish between the intraband impurity scattering, which does not add any new physics (in the Born approximation) compared with single-band superconductivity, and \textit{interband} scattering, which in many cases has an effect comparable to the pair-breaking effect of magnetic impurities (or of nonmagnetic impurities in superconductors with $p$- or $d$-wave pairing)~\cite{PhysRevB.55.15146}. In this regard, the fact that no strong correlation has been observed between the residual resistivity
(which indirectly characterizes the impurity scattering) and the critical
temperature $T_{c}$ of the (nearly) optimally electron-doped BKFA (see Table~\ref{table:residual}) indicates that the level of \textit{interband} impurities in the Born limit is very small. Thus one only needs to estimate the intraband scattering rates $\gamma_{\mathrm{A}},\ \gamma_{\mathrm{B}}$.
\begin{table}
\caption{\label{table:residual}Superconducting transition temperature and residual resistivity $\rho_{40\textrm{K}}$ of (nearly) optimally hole-doped BKFA.}
\begin{tabular}{c c c}
\hline
$T_{\textrm{c}}$, K&Residual resistivity, $\textrm{m}\Omega\ \textrm{cm}$&Reference\\
\hline
38.5&0.04&\cite{PhysRevLett.105.027003s}\\
38&0.075&\cite{PhysRevB.78.224512}\\
38&0.1&\cite{PhysRevLett.101.107006}\\
36.5&0.055&\cite{PhysRevB.79.174501}\\
\hline
\end{tabular}
\end{table}
\subsection{Theoretical model}
As a starting point we consider a $4-$band model based on the band-structure calculations with two hole bands and two electron bands crossing the Fermi level that has proven successful in accounting for the thermodynamical properties of BKFA~\cite{PhysRevLett.105.027003s}. We use the same input parameters, namely, the densities of states $N_1=22\ \textrm{Ry-st}^{-1}$, $N_2=25\ \textrm{Ry-st}^{-1}$, and $N_3=N_4=7\ \textrm{Ry-st}^{-1}$, the first two having a hole while the other two an electron character. The main input, the spectral function of the intermediate boson, was taken following Ref.~S\onlinecite{PhysRevB.78.134524} in the form of a spin-fluctuation spectrum $\tilde{B}_{ij}(\Omega )=\lambda _{ij}f(\Omega /\Omega _{sf})$ with a linear $\omega$ dependence at low frequencies. Here $\lambda_{ij}$ is the coupling constant pairing band $i$ with band $j$ and $\Omega _{SF}$ is a characteristic spin-fluctuation frequency, the values of which correspond to those in Ref.~\onlinecite{PhysRevLett.105.027003s}: $\Omega_{sf}=13\ \textrm{meV}$ and%
\begin{equation}
\lambda _{ij}=\left( 
\begin{array}{cccc}
0.2 & 0 & -1.7 & -1.7 \\ 
0 & 0.2 & -0.25 & -0.25 \\ 
-5.34 & -0.89 & 0.2 & 0 \\ 
-5.34 & -0.89 & 0 & 0.2%
\end{array}%
\right).\label{l4x4}
\end{equation}%
Negative elements correspond to \textit{interband} hole-electron
repulsion, while the positive --- to \textit{intraband}
attraction.

In order to apply this full 4-band model to description of the transport properties one has to take into account an additional set of 4 plasma frequencies and a 4x4 matrix of impurity scattering rates. The latter are difficult to determine theoretically and thus would have to be treated as free paremeters of the model. It would render the problem highly overparametrized. On the other hand, in the case of BKFA it is known that three larger gaps have approximately the same value $|\Delta_{\mathrm{A}}|\equiv|\Delta_1|\approx|\Delta_3|\approx|\Delta_4|\approx9\ \textrm{meV}$, while the smaller gap is $|\Delta_{\mathrm{B}}|\equiv|\Delta_2|\approx3\ \textrm{meV}$. One can assume this restriction exact and introduce it into the theory thus reducing the original 4-band model to a more tractable 2-band model as explained in the following section.

\subsection{Reduction to a two-band model}

In general, the superconducting order parameters are a solution of a linear system of equations
\begin{equation}
e_{i}=\sum_{j=1}^{4}B_{ij}(\omega)e_{j}.\label{eq:gapsystem}
\end{equation}
The Eliashberg functions $B_{ij}(\omega)$ satisfy the symmetry relations
\begin{equation}
N_iB_{ij}(\omega )=N_jB_{ji}(\omega)\label{eq:symmetryrelations}
\end{equation}%
and, therefore, can be represented in the form $B_{ij}(\omega)=U_{ij}(\omega)N_j$, where $U_{ij}$ is a symmetrical matrix. Further, we can construct a functional
\begin{equation}
{\mathfrak{F}}\{e_i\}=\sum_{j=1}^{4}N_je_j^2-\sum_{i,j=1}^4N_ie_iU_{ij}N_je_j.  \label{eq:functional}
\end{equation}%
Equation~(\ref{eq:gapsystem}) then results from minimization of ${\mathfrak{F}}$ with respect to $e_i$. As mentioned above, BKFA has three gaps with very
close absolute values (the first hole gap has the opposite sign with respect to the other two). 
\begin{figure}[!b]
\href{http://www.fkf.mpg.de/keimer/groups/optical/bkfafir.jar}{\includegraphics[width=\columnwidth]{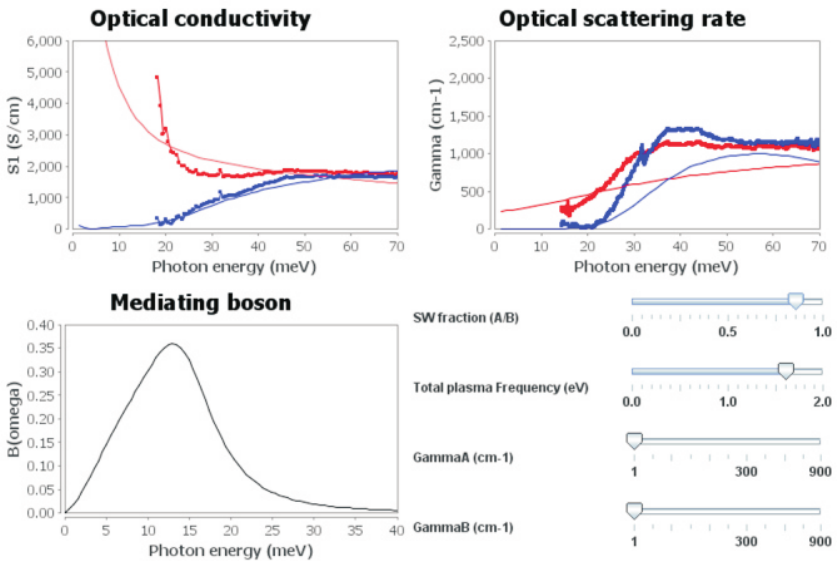}}
\caption{\label{fig:interactive}(INTERACTIVE: click on the image to run the simulation)~Real part of the optical conductivity (top left) and the optical scattering rate (top right) of $\textrm{Ba}_{0.68}\textrm{K}_{0.32}\textrm{Fe}_2\textrm{As}_2$ in the superconducting state at $10\ \textrm{K}$ (blue lines) and the normal state at $40\ \textrm{K}$ (red) as obtained experimentally (heavy) and from the effective two-band Eliashberg theory (thin). (bottom left) The spin-fluctuation spectrum used in the simulation in both the superconducting and the normal state. (bottom right) Interactive controls: adjust the sliders to select the physical parameters of the model.}
\end{figure}
Minimizing functional Eq.~(\ref{eq:functional}) subject to the additional constraints $e_3=e_4=-e_1=\Delta
_{\mathrm{A}}$, and $e_2=-\Delta _{\mathrm{B}}$ one finds
\begin{equation}
\left( 
\begin{array}{c}
\Delta _{\mathrm{A}} \\ 
\Delta _{\mathrm{B}}%
\end{array}%
\right) =\left( 
\begin{array}{cc}
\lambda_{AA}&\lambda_{AB}\\ 
\lambda_{BA}&\lambda_{BB}%
\end{array}%
\right)\left( 
\begin{array}{c}
\Delta_{\mathrm{A}} \\ 
\Delta_{\mathrm{B}}%
\end{array}%
\right),
\end{equation}%
where the matrix elements satisfy 
\begin{eqnarray*}
\lambda_{AA}&=&\frac{N_1\left( \lambda_{11}-2\lambda_{13}-2\lambda_{14}\right) +N_3\lambda_{33}{+}N_4\lambda_44}{N_1+N_3+N_4}, \\
\lambda_{AB} &=&\frac{N_2\left(\lambda_{23}+\lambda_{24}\right)}{%
N_1+N_3+N_4}, \\
\lambda_{BA}&=&\lambda_{23}+\lambda_{24}, \\
\lambda_{BB}&=&\lambda_{22}.
\end{eqnarray*}%
Assuming the matrix elements Eq.~(\ref{l4x4}), the following coupling constants of the reduced 2-band model are obtained:
\begin{equation}
\lambda_{IJ}=\left( 
\begin{array}{cc}
4.36&-0.35 \\ 
-0.5&0.2%
\end{array}%
\right) ,\text{ }I,J=\{A,B\}.\label{l2x2}
\end{equation}%
The partial densities of states on the Fermi level of effective band $A$ and band $B$ are%
\begin{eqnarray}
N_{\mathrm{A}}&=&N_1+N_3+N_4=36\ \textrm{Ry-st}^{-1},\nonumber\\ N_{\mathrm{B}}&=&N_2=25\ \textrm{Ry-st}^{-1}\label{dos2b}.
\end{eqnarray}%
Interestingly, even though the main interactions in the 4-band
model with the coupling constants Eq.~(\ref{l4x4}) come from the nondiagonal elements, in the reduced 2-band counterpart they are incorporated into the effective \textit{intraband} $\lambda_{\mathrm{AA}}$ matrix element. Figure~\ref{fig:interactive} presents an interactive simulation of this effective two-band model (click on the figure to run the simulation and adjust the sliders to set the physical parameters of the system).
\subsection{Verification of the $2\times 2$ model}
\begin{figure}[!t]
\includegraphics[width=\columnwidth]{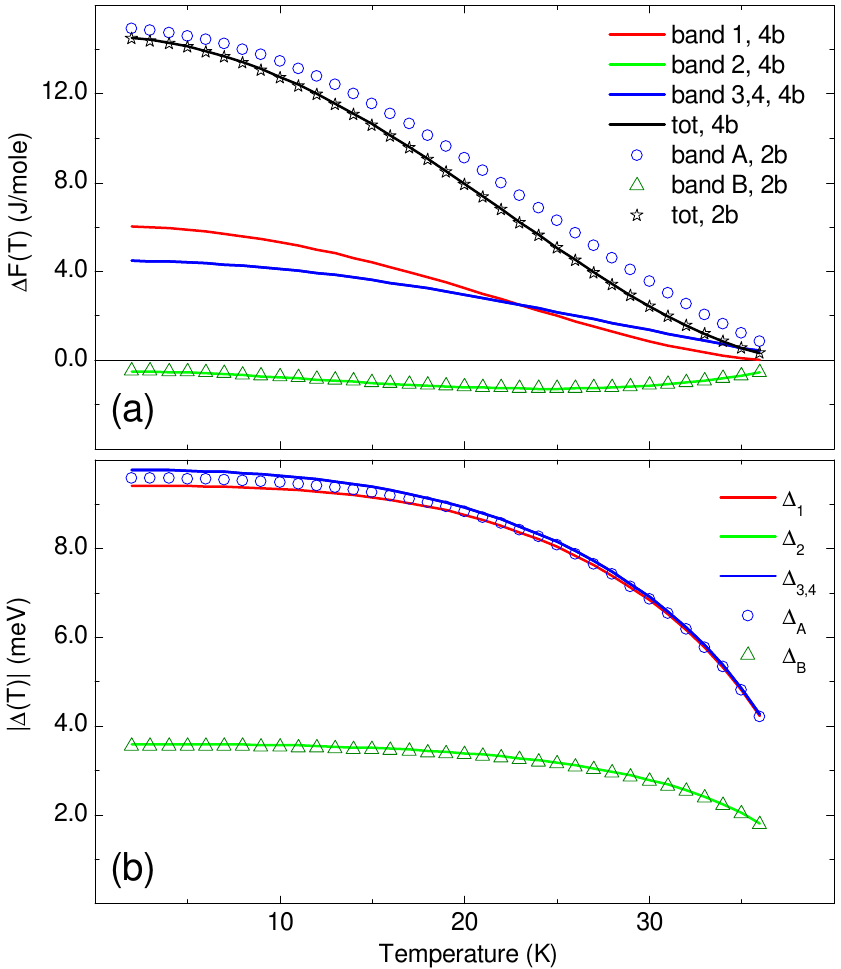}
\caption{\label{fig:verification}Temperature depencence of the (a)~free energy and (b)~superconducting gaps the 4-band (lines) and reduced 2-band (symbol) models, with coupling matrices Eqs.~(\ref{l4x4},\ref{l2x2}), respectively.}
\end{figure}

The effective 2-band model closely reproduces all the predictions of the 4-band model such as the superconducting transition temperature $T_{\mathrm{c}}=38.4\ \textrm{K}$, superconducting gaps $\Delta_{\mathrm{A}}=9.7\ \textrm{meV}$ and $\Delta_{\mathrm{B}}=3.7\ \textrm{meV}$, free energy and superconducting gaps as functions of temperature, as shown in Figs.~\ref{fig:verification}(a) and~(b), respectively. The calculated densities of states $N_{\mathrm{A}}$ and $N_{\mathrm{B}}$ are very similar, in accordance with the partial Sommerfeld constants obtained in the treatment of the specific heat data in a phenomenological two-band $\alpha$-model~\cite{PhysRevLett.105.027003s}: $\gamma_{\mathrm{A}}\simeq\gamma_{\mathrm{B}}$ (the Sommerfeld constant $\gamma$ is related to the density of states $N$ via $\gamma =\frac{2\pi}{3}N$).

\clearpage
\end{document}